\documentstyle[multicol,floats,epsfig,aps,prl]{revtex}
\begin{document}
\draft
\title{Governing dynamics by squeezing in a system of cold 
trapped ions}
\author{G. Ramon, \cite{email1} C. Brif, \cite{email2} 
and A. Mann \cite{email3}}
\address{Department of Physics, Technion -- Israel Institute of 
Technology, Haifa 32000, Israel}
\maketitle

\begin{abstract}
We consider a system of laser-cooled ions in a linear harmonic 
trap and study the phenomenon of squeezing exchange between
their internal and motional degrees of freedom. An interesting 
relation between the quantum noise reduction (squeezing) and 
the dynamical evolution is found when the internal and motional 
subsystems are prepared in properly squeezed (intelligent) 
states. Specifically, the evolution of the system is fully 
governed by the relative strengths of spectroscopic and motional 
squeezing, including the phenomenon of total cancellation of the 
interaction when the initial squeezing parameters are equal.
\end{abstract}


\begin{multicols}{2}
In the last few years significant progress has been achieved, 
both theoretically and experimentally, in the coherent manipulation 
of quantum states of trapped atomic ions \cite{Win:rev}. 
Attention is mainly devoted to three mutually related problems: 
(i) generation and analysis of non-classical states of motion
\cite{MS:exp,MS:rev}, 
(ii) implementation of quantum logic and computation
\cite{QLogic}, 
and (iii) generation of entangled states which can improve 
signal-to-noise ratio in spectroscopy \cite{Win:spectr,AgPu94}. 
In this work we discuss the exchange of squeezing between 
the internal and motional degrees of freedom of a system of cold 
trapped ions. This discussion will reveal a fundamental relation 
that exists between the quantum noise reduction (squeezing) in 
this important physical system and its dynamical properties.
Our main result is that under specific initial conditions, when 
the internal and motional subsystems are prepared in properly 
squeezed (intelligent) states, squeezing properties of the 
system govern its dynamical evolution.

It is known that the quantum noise in spectroscopy can be 
reduced using the so-called squeezed atomic states 
\cite{Win:spectr,AgPu94}. 
In Ramsey spectroscopy \cite{Ramsey} one deals with $N$ two-level 
systems (e.g., trapped atomic ions) interacting with classical 
radiation fields. One can equivalently describe this physical 
situation as the interaction of $N$ spin-half particles with 
classical magnetic fields. Denoting by ${\bf S}_i$ the spin of 
the $i$th particle, one can use the collective spin operators, 
${\bf J} = \sum_{i} {\bf S}_i$. The basic set of states is 
$|j,m\rangle$ ($m=j,j-1,\ldots,-j$):
$$
{\bf J}^2 |j,m\rangle = j(j+1) |j,m\rangle ,
\hspace{5mm} J_z |j,m\rangle = m |j,m\rangle ,
$$
and $j=N/2$ if only the symmetric Dicke states $|j,m\rangle$ 
are considered \cite{Dicke54}.
A spectroscopic (or interferometric) process can be described 
mathematically in the Heisenberg picture as a unitary 
transformation 
\begin{equation}
{\bf J}_{{\rm out}} = U(\phi) {\bf J} U^{\dagger}(\phi)
= {\sf U}(\phi) {\bf J},
\end{equation}
where ${\sf U}(\phi)$ is a $3 \times 3$ transformation matrix 
and $\phi $ is the parameter to be estimated. 
The standard transformation consists of three steps: rotation 
around the $y$ axis by $\pi /2$, rotation around the 
$z$ axis by $\phi $, and rotation around the $y$ axis 
by $- \pi /2$. In the Ramsey method the rotations 
are performed through the application of magnetic fields of the 
type ${\bf B} = B \hat{{\bf y}} \sin \omega t$. The phase shift
is $\phi = ( \omega - \omega _0 ) T$, where $\omega_0$ is the 
frequency of the transition between the atomic levels and $T$ 
is the duration of the second Ramsey pulse. 
After the transformation ${\sf U}(\phi)$, one measures the 
population inversion represented by the operator 
$J_{z\, {\rm out}} = \cos \phi J_z - \sin \phi J_y$, and the 
information about $\phi$ is inferred from the value 
$\langle J_{z\, {\rm out}} \rangle $. The uncertainty 
of the frequency measurement is 
\begin{equation}
\delta \omega_0 = \frac{\Delta J_{z\, {\rm out}}}{|\partial 
\langle J_{z\, {\rm out}}\rangle /\partial \omega _0|}.  
\label{dw1}
\end{equation}
Taking for simplicity $\phi =\pi /2$, this gives 
\begin{equation}
\delta \omega_0 = \frac{\Delta J_y}{T|\langle J_z\rangle |}.  
\label{dw2}
\end{equation}

If one uses a Dicke state $|j,m\rangle $ at the input, then 
$\delta \omega_0 = (T|m|)^{-1} 
\left\{ \frac{1}{2} [j(j+1)-m^{2}] \right\}^{1/2}$. 
This uncertainty is minimized for $m=\pm j$: 
\begin{equation}
(\delta \omega_0)_{{\rm SNL}} = 1/( T \sqrt{N} ) ,
\end{equation}
which is the shot-noise limit (SNL). The measure of spectroscopic 
squeezing is the parameter 
\begin{equation}
\xi_R = \frac{\delta \omega_0}{(\delta \omega_0)_{{\rm SNL}}}
= \sqrt{N} \frac{\Delta J_y}{|\langle J_z \rangle |}.  \label{xir}
\end{equation}
A quantum state for which $\xi _R<1$ is spectroscopically 
squeezed, and the use of such a state will reduce the quantum 
noise of a spectroscopic measurement below the shot-noise limit. 
The Heisenberg limit of $\delta \omega_0$ is 
$[ T \sqrt{2j(j+1)} ]^{-1}$, that is $\xi _R$ = $(j+1)^{-1/2}$,
which is obtained from the uncertainty relation
\begin{equation}
\Delta J_x \Delta J_y \geq \frac{1}{2} |\langle J_z \rangle| 
\label{Hur}
\end{equation}
and the condition $( \Delta J_x )^2 \leq \frac{1}{2} j(j+1)$.

It can be shown that spectroscopic squeezing is achieved with 
the so-called intelligent (minimum-uncertainty) states 
\cite{Win:spectr,AgPu94,IS:orig,IS:gen,IS:interf}. 
The $J_x$-$J_y$ intelligent states are 
defined as states which give the equality in the uncertainty 
relation (\ref{Hur}). They are determined by the eigenvalue 
equation \cite{IS:orig,IS:gen}
\begin{equation}
( \eta J_x - i J_y )|\lambda,\eta\rangle 
= \lambda |\lambda,\eta\rangle ,
\label{isee}
\end{equation}
where $\lambda $ is a complex eigenvalue and $\eta $ is a real 
parameter given by $|\eta |=\Delta J_y/\Delta J_x$. 
For $|\eta |<1$, the intelligent states are squeezed in $J_y$ and 
anti-squeezed in $J_x$, which results in spectroscopic squeezing. 
Equation (\ref{isee}) can be solved using the analytic 
representation method \cite{Brif}. The spectrum is
discrete: $\lambda =im_0\sqrt{1-\eta ^2}$ where 
$m_0=j,j-1,\ldots ,-j$. The intelligent states can be written
in the Dicke basis using the Jacobi polynomials 
$P_n^{(\alpha ,\beta )}(x)$ \cite{IS:interf,Brif}. The expectation
value of $J_z$ and the squeezing parameter for the intelligent 
states are 
\begin{eqnarray}
& \langle J_z \rangle = - j \eta {\cal F} , \hspace{8mm}
\xi_R = {\cal F}^{-1/2} , & \\
& {\cal F} = \displaystyle{ 1 + \frac{j+|m_{0}|}{j}
(1-\eta^{2}) \frac{ P_{j-|m_{0}|-1}^{(1,-2j)}(1-2\eta^{2}) }{
P_{j-|m_{0}|}^{(0,-2j-1)}(1-2\eta^{2}) } }. &
\end{eqnarray}
In the range $|\eta |<1$ we always find $\xi_R \leq 1$. 
For $m_0=\pm j$ the intelligent states are simultaneously the 
atomic coherent states with $\xi_R = 1$. Numerical results show
that the minimum value of $\xi_R$ (the best value of squeezing) 
for given $j$ and $\eta$ is achieved for the intelligent states 
with $m_{0}=0$. For these states the Heisenberg limit is achieved 
as $|\eta| \rightarrow 0$.

Wineland {\em et al.} \cite{Win:spectr} have studied the 
possibility to generate spectroscopic squeezing using an 
interaction that couples the internal electronic levels
of laser-cooled ions and a center-of-mass mode of their 
quantized oscillatory motion in a linear harmonic trap. 
The interaction Hamiltonian is ($\hbar = 1$)
\begin{equation}
H = g ( a^{\dagger} J_{-} + a J_{+} ),  \label{DHam}
\end{equation}
where $g$ is the coupling constant, $a$ and $a^{\dagger}$ 
are the boson operators of the center-of-mass mode of the 
oscillatory motion, and $J_{\pm} = J_x \pm i J_y$ are the 
collective ionic raising and lowering operators. 
The Hamiltonian (\ref{DHam}) is the same as in the 
Tavis-Cummings version \cite{TC68} of the Dicke model 
\cite{Dicke54} in quantum optics, in which a collection of 
two-level atoms interacts with a single-mode radiation 
field inside a cavity. This model can be considered as a 
multiatom generalisation of the Jaynes-Cummings model 
\cite{JCM}.
An important feature of the realization with cold trapped 
ions is that decoherence processes can be made very small 
during the interaction.

The Hilbert space ${\cal H}$ of the whole system can be 
decomposed into a direct sum of finite-dimensional 
invariant subspaces ${\cal H}_L$: 
\begin{equation}
{\cal H} = \bigoplus_{L=0}^\infty {\cal H}_L , \label{dec}
\end{equation}
where $L=a^{\dagger }a+J_z+N/2$ is the total excitation 
(a constant of motion). 
Each subspace ${\cal H}_L$ is spanned by the orthonormal basis 
$|j,L-j-n\rangle_{{\rm ion}} \otimes |n\rangle_{{\rm osc}}$, 
where $|n\rangle_{{\rm osc}}$ are the Fock states of the 
oscillator. 
If the oscillator is initially in a Fock state and the 
ions are in a Dicke state, the system will evolve in one 
invariant subspace. For the oscillator and/or ions prepared 
initially in a superposition state, one should take into 
account contributions from different subspaces. 
The exact solution of the problem is obtained by  
diagonalization of the interaction Hamiltonian (\ref{DHam}) 
in each of the invariant subspaces ${\cal H}_L$ involved 
\cite{TC68}.

Wineland {\em et al.} \cite{Win:spectr} considered the 
situation when the ions are prepared initially in the 
internal ground state, 
$|j,-j\rangle_{{\rm ion}} = \prod_{i} |-\rangle_i$,
while the oscillator is prepared in the squeezed vacuum state 
$|\xi_q \rangle_{{\rm osc}}
= \sum_{n} b_n |n\rangle_{{\rm osc}}$, where
\begin{eqnarray}
&& b_n = \frac{\sqrt{n!}}{(n/2)!} \sqrt{
\frac{ 2\xi_q }{1+ \xi_{q}^{2} } }
\left( -\frac{1}{2} \frac{ 1-\xi_{q}^{2} }{ 1+\xi_{q}^{2} }
\right) ^{n/2}
\hspace{2mm} {\rm for} \;\; n \;\; {\rm even},  \nonumber \\
&& b_n = 0 \hspace{10mm} {\rm for} \;\;n\;\; {\rm odd}.  
\label{bn}
\end{eqnarray}
Here, $q = (a+a^{\dagger })/\sqrt{2}$ is the position and 
\begin{equation}
\xi_q = \Delta q / (\Delta q)_{{\rm vac}} = \sqrt{2} \Delta q
\end{equation}
is the squeezing parameter of the oscillator. Squeezing in $q$ 
is obtained if $\xi_q < 1$. Numerical study shows 
that motional squeezing can be transferred during the 
interaction into spectroscopic squeezing of the ions. The minimal 
value $(\xi_R)_m$ achieved during the interaction depends on the 
initial value $\xi_q(0)$ of motional squeezing.

Let us address the following question: How will the system evolve
in the Dicke-model interaction if the ions are prepared in a 
correlated state with initial spectroscopic squeezing? 
We start by considering an even number of ions prepared in the 
spectroscopic intelligent state $|\eta\rangle_{{\rm ion}} = 
\sum_{m} c_m |j,m\rangle_{{\rm ion}}$ with $m_0 = 0$. Here  
\begin{equation}
c_{-j+2r} = {j \choose r} {2j \choose 2r}^{-1/2}
\left( \frac{1-\eta}{1+\eta} \right)^r c_{-j} ,  \label{cm}
\end{equation}
$r=0,1,\ldots,j$, and $c_{-j}$ is determined by the 
normalization.
The oscillator is prepared in the squeezed vacuum state, 
so the initial wave function of the whole system is
\begin{equation}
|\psi(0)\rangle = |\eta\rangle_{{\rm ion}} \otimes 
|\xi_q \rangle_{{\rm osc}} .
\label{eig1}
\end{equation}

Letting the state (\ref{eig1}) evolve in time, governed by the 
interaction Hamiltonian (\ref{DHam}), one can calculate the 
time-dependent reduced density matrix of the ionic internal 
subsystem by tracing out the motional (oscillator) 
degree of freedom:
\begin{equation}
\rho_{\rm{ion}}( \tau ) = {\rm Tr}_{\rm{osc}}
\left( | \psi ( \tau ) \rangle \langle \psi ( \tau ) | \right) .
\end{equation}
Here $\tau = g t$ is the scaled time.
Then one can calculate the von Neumann entropy 
$S = - {\rm Tr}\, ( \rho \ln \rho )$ 
for the internal subsystem and evaluate from it the degree 
of entanglement between the internal and motional subsystems. 
Since both subsystems are initially in pure states, 
the entropy of the whole system is always zero. 
Therefore, the marginal entropies of the two subsystems are 
equal, $S_{\rm{osc}}( \tau ) = S_{\rm{ion}}( \tau )$. 

As the system evolves, the marginal entropy 
$S_{\rm{ion}}(\tau)$ rapidly oscillates; the time average 
$\overline{ S_{\rm{ion}} }$ and the oscillation amplitude
$(S_{{\rm ion}})_{{\rm max}} - (S_{{\rm ion}})_{{\rm min}}$ 
depend on the initial state parameters $\eta$ and $\xi_{q}(0)$.
Figure 1 shows $\overline{ S_{\rm{ion}} }$ and 
$(S_{{\rm ion}})_{{\rm max}} - (S_{{\rm ion}})_{{\rm min}}$
as functions of $\eta$, for two ions and $\xi_{q}(0) = 0.6$. 
One can see that as $\eta$ approaches the value $0.36$, both 
the average and the oscillation amplitude of $S_{\rm{ion}}$ 
rapidly decrease. At the point $\eta = 0.36$ the entropy 
$S_{\rm{ion}}(\tau)$ remains zero for any time, implying that 
the two subsystems are permanently disentangled. Actually, we 
find that for $\eta = \xi_{q}^{2}(0)$ the initial state of the 
system remains unchanged in time, as can be verified by 
examining the time evolution of the density matrix. 
So, an equilibrium point exists in the parameter space, for 
which the interaction between the subsystems is effectively 
cancelled. Of course, this situation is possible only if the 
initial state is an eigenstate of the interaction Hamiltonian. 

Let us study the phenomenon of the interaction cancellation
in a more general way. Any product state of the form 
\begin{equation}
|\psi\rangle = \sum_{m=-j}^{j} \tilde{c}_m 
|j,m\rangle_{\rm{ion}} \otimes 
\sum_{n=0}^{\infty}  \tilde{b}_n |n\rangle_{\rm{osc}}
\label{prodstate}
\end{equation}
can be decomposed into a sum over the subspaces ${\cal H}_L$:
\begin{eqnarray*}
& & |\psi\rangle = \sum_{L=0}^{\infty} 
|\psi_{L}\rangle , \\
& & |\psi_{L}\rangle = \sum_{s=0}^{{\rm min}(N,L)} 
\tilde{c}_{-j+s} \tilde{b}_{L-s}\, |j,-j+s\rangle_{\rm{ion}} 
\otimes |L-s\rangle_{\rm{osc}} , 
\end{eqnarray*}
The state $|\psi\rangle$ can be an eigenstate of the
Hamiltonian if and only if all $|\psi_{L}\rangle$
are eigenstates with the same eigenvalue independent of $L$.
The only eigenvalue of the Hamiltonian (\ref{DHam}) that 
appears in different subspaces ${\cal H}_L$ is zero (this 
eigenvalue actually appears for any odd-dimensional subspace) 
\cite{TC68}. Then we are able to prove that the product 
state of the form (\ref{prodstate}) is an eigenstate
of the Hamiltonian (\ref{DHam}), if and only if it is the 
state $|\eta\rangle_{{\rm ion}} \otimes 
|\xi_q \rangle_{{\rm osc}}$ of Eq.\ (\ref{eig1}) with 
\begin{equation}
\eta = \tilde{\eta} ,
\label{cond1}
\end{equation}
where $\tilde{\eta} = \xi_{q}^{2}$. 
At this point it is worth noting that the squeezed vacuum state 
is an intelligent state of the Weyl-Heisenberg group, obeying the
eigenvalue equation
\begin{equation}
(\tilde{\eta} p -i q) |\xi_{q}\rangle = 0 .
\label{sveq}
\end{equation}
Here $p = (a-a^{\dagger})/ i \sqrt{2}$ is the momentum of the 
oscillator, and $\tilde{\eta}$ is a positive parameter given by 
$\tilde{\eta} = \xi_{q}^{2} = \Delta q/\Delta p$. 
Equation (\ref{sveq}) is similar to the equation
$(\eta J_x -i J_y ) |\eta\rangle = 0$ satisfied by the 
spectroscopic intelligent state $|\eta\rangle$ with $m_{0} = 0$.
For this reason we call the product state (\ref{eig1}) 
the ``double intelligent'' state. It is interesting that 
the condition (\ref{cond1}) can be rewritten in the form
\begin{equation}
\frac{\Delta q}{\Delta p} = \frac{\Delta J_{y}}{\Delta J_{x}} , 
\label{eigcond1}
\end{equation}
which implies that the interaction between the internal
and motional subsystems is cancelled when they are equally
squeezed (or ``equally intelligent''). In other words, the
dynamical equilibrium of the system is determined by 
the squeezing equilibrium between its subsystems.

Next we consider what happens if one prepares the ions in the 
``double intelligent'' state of Eq.\ (\ref{eig1}) with the 
squeezing parameters slightly detuned from the 
equilibrium point. For example, we can make a perturbative 
expansion of the time-dependent state around the equilibrium 
point, with $\eta - \xi_{q}^{2}(0)$ being the small parameter. 
Then it becomes apparent that as $\eta$ gets closer to 
$\xi_{q}^{2}(0)$, the state becomes more similar to the 
Hamiltonian eigenstate and its change with time is less 
pronounced. Also, the deviation of the marginal entropy
$S_{{\rm ion}}(\tau)$ from its initial value (i.e., from zero) 
rapidly decreases as the squeezing parameters approach the 
equilibrium point (see Fig.~1). Therefore, we can conclude 
that the equilibrium point is stable. 

Figure 2 reveals another interesting feature of the 
``double intelligent'' initial state (\ref{eig1}), namely, the 
exchange of squeezing between the motional and internal degrees 
of freedom that occurs during the interaction. For values 
of $\xi_q(0)$ above the equilibrium point, spectroscopic 
squeezing deteriorates [the time average and minimal values  
$\overline{ \xi_R }$ and $( \xi_R )_m$ are greater than the
initial value $\xi_R (0)$], while motional squeezing is 
improved [the time average and minimal values  
$\overline{ \xi_q }$ and $( \xi_q )_m$ are less than the
initial value $\xi_q (0)$]. Correspondingly, for $\xi_q(0)$ 
below the equilibrium point, spectroscopic squeezing is improved
[except for very low values of $\xi_q (0)$], while motional 
squeezing deteriorates. Thus the exchange of spectroscopic and
motional squeezing occurs, depending on their initial strengths.
Figure 2 also shows the deviation of the time average 
$\overline{\langle J_z \rangle}$ of the population inversion
from its initial value $\langle J_z \rangle (0)$. Similarly
to the behavior of the squeezing parameters, the deviation
$\overline{\langle J_z \rangle} - \langle J_z \rangle (0)$
changes its sign at the equilibrium point. Thus the behavior
of the system prepared in the ``double intelligent'' state
is governed by the relative strengths of spectroscopic and 
motional squeezing.

The phenomenon in which the dynamics of the system is governed 
by its squeezing properties appears exclusively for the 
intelligent spectroscopic state with $m_0=0$ interacting 
with the squeezed vacuum. (Note that the condition $m_0=0$
can be satisfied only for integer $j$, i.e., for even numbers
of ions.) If one prepares, for example, the internal 
subsystem in a spectroscopic intelligent state with $m_0\neq 0$,
no equilibrium point will appear since such a state cannot be an 
eigenstate of the interaction Hamiltonian. Moreover, numerical 
calculations show that no exchange of squeezing between the 
internal and motional subsystems occurs in the entire parameter
space. Instead, squeezing deteriorates with time for both 
subsystems.

The generation of the spectroscopic intelligent states is still 
an open question. A number of schemes were proposed that employ 
the interaction of atoms (ions) with the broadband squeezed 
vacuum \cite{AgPu90}. 
Alternatively, it is known \cite{Win:rev,QLogic} that 
quantum logic schemes can be used to generate various correlated 
states of the cold trapped ions. In particular, the quantum logic 
methods are suitable for producing the spectroscopic intelligent 
state $|\eta\rangle_{{\rm ion}}$ for a pair of ions. In this case 
we obtain $|\eta\rangle_{{\rm ion}} = 
\sin\theta |1,-1\rangle + \cos\theta |1,1\rangle$, 
where $\theta =\tan^{-1} \left[ (1+\eta)/(1-\eta) \right]$
and $\xi_R = \left[ \frac{1}{2} ( 1+\eta^2 ) \right]^{1/2} 
= \left( 1 + \sin 2\theta \right)^{-1/2}$.
The equilibrium condition (\ref{eigcond1}) can be written as  
$\xi_{q}^{2} = \tan \left( \theta -\pi /4\right)$. 

In conclusion, we discussed interesting quantum phenomena that 
take place for a system of cold trapped ions prepared in the
``double intelligent'' state. The uniqueness of this state is 
that squeezing properties of the system govern its dynamical 
evolution. In particular, when the squeezing strengths of the 
internal and motional subsystems are equal, the system is in 
the state of stable dynamical equilibrium. Around the
equilibrium point exchange of squeezing between the
subsystems occurs.

This work was supported by GIF --- German-Israeli Foundation
for Research and Development, by the Fund for Promotion of 
Research at the Technion, and by the Technion VPR Fund --- 
Promotion of Sponsored Research.
G.R. and C.B. gratefully acknowledge the financial help from 
the Technion. 


\end{multicols}

\twocolumn
\begin{figure}[htbp]
\epsfxsize=0.45\textwidth
\centerline{\epsffile{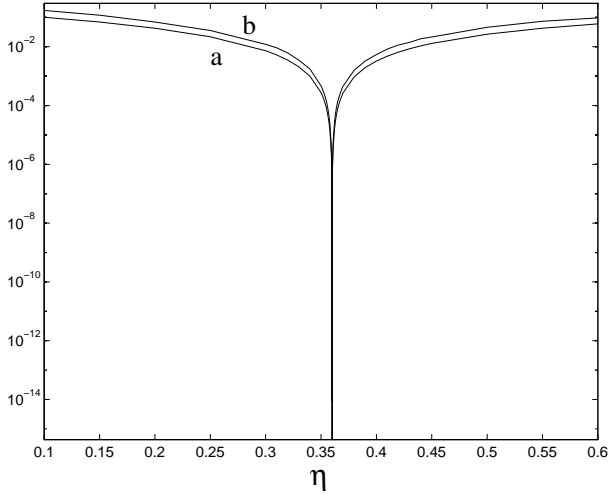}}
\vspace*{1mm}
\caption{Two ions prepared in the state 
$|\eta\rangle_{{\rm ion}} \otimes |\xi_q \rangle_{{\rm osc}}$
with $\xi_q (0) = 0.6$.
(a) The time average $\overline{S_{{\rm ion}}}$ 
and (b) the oscillation amplitude 
$(S_{{\rm ion}})_{{\rm max}} - (S_{{\rm ion}})_{{\rm min}}$ 
of the marginal entropy of the internal subsystem versus
$\eta$.}
\end{figure}
\begin{figure}[htbp]
\epsfxsize=0.45\textwidth
\centerline{\epsffile{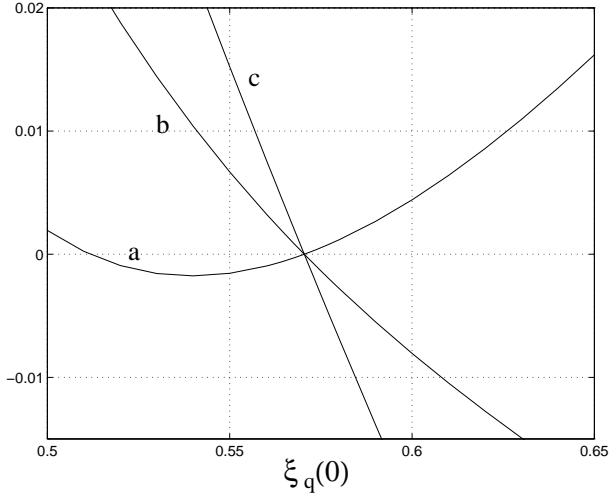}}
\vspace*{1mm}
\caption{Two ions prepared in the state 
$|\eta\rangle_{{\rm ion}} \otimes |\xi_q \rangle_{{\rm osc}}$
with $\eta = 0.3254$. The time average deviations of 
(a) the spectroscopic squeezing parameter, 
$\overline{ \xi_R } - \xi_R (0)$, 
(b) the motional squeezing parameter,
$\overline{ \xi_q } - \xi_q (0)$, and
(c) the population inversion,
$\overline{ \langle J_z \rangle } - \langle J_z \rangle (0)$,
plotted versus $\xi_q (0)$.}
\end{figure}

\end{document}